%% file: main.tex
\documentclass[a4paper, oneside, twocolumn, notitlepage, 10pt]{extarticle_ecoc}
\usepackage{ecoc}

\begin{document}
\selectlanguage{english}    


\title{Measurement and Analysis of the Power Consumption of Hybrid-Amplified SCL-band Links}%


\author{
    Ronit~Sohanpal\textsuperscript{(1)}, Jiaqian~Yang\textsuperscript{(1)}, Eric~Sillekens\textsuperscript{(1)}, Henrique~Buglia\textsuperscript{(2)}, \\Mingming~Tan\textsuperscript{(3)}, 
    Dini~Pratiwi\textsuperscript{(3)}, Robert~I.~Killey\textsuperscript{(1)}, Polina~Bayvel\textsuperscript{(1)}
}

\maketitle                  


\begin{strip}
    \begin{author_descr}

        \textsuperscript{(1)} Optical Networks Group, UCL (University College London), London, UK, 
        \textcolor{blue}{\uline{ronit.sohanpal@ucl.ac.uk}}\\
        \textsuperscript{(2)} Nokia, San Jose, CA, USA\\
        \textsuperscript{(3)} Aston Institute of Photonic Technologies, Aston University, Birmingham, B4 7ET, UK\\
    \end{author_descr}
\end{strip}

\renewcommand\footnotemark{}
\renewcommand\footnoterule{}


\begin{strip}
    \begin{ecoc_abstract}
        We studied the power consumption of hybrid-amplified SCL-band links using commercial benchtop amplifiers and Raman pumps. We show a reduction in energy per bit for multi-span hybrid Raman amplified links of up to 26\% versus lumped amplification.
        \textcopyright2025 The Author(s)
    \end{ecoc_abstract}
\end{strip}


\section{Introduction} \vspace{-0.3em}
To meet increasing capacity demands, the last decade has seen a strong research push towards long-haul DWDM coherent transmission systems beyond the C- and L- wavelength bands. A variety of different UWB schemes have been proposed, ranging from adding just the S-band to occupying the entire OESCLU spectrum and even extending further towards 2~\textmu m wavelengths with the newly-defined X-band \cite{hamaoka2024_110,hamaoka2025_107,yang2024experimental,yang2025transmission,aparecido2024experimental,zhang2024_201,zhang2025_214,kobayashi2024c,puttnam400,shimizu2024_133,shimizu2025_27}. 

In parallel to this capacity-motivated research there has been a strong push to reduce the power consumption of coherent links \cite{pillai2014end,sinkin2022strategies,Downie2022}. Alongside ecological and financial concerns, a major motivator for subsea links is the limited electrical power delivery to the repeaters, which constrains the design and performance of the link \cite{frisch2013electrical,desbruslais2015maximizing,maher2016capacity}. Currently, only one C+L-band subsea link has seen deployment, with the rest being C-band only \cite{plcn}. 

Further increasing link throughput by implementing UWB schemes raises several questions about the amplification power efficiency. In C-band systems alone, the link amplifiers can contribute as much as 18\% of the total power consumption \cite{pillai2014end}. Since each distinct wavelength band requires its own tailored amplifier, the addition of other wavelength bands leads to rapid scaling of amplifier power requirements. In addition, many of the amplifiers for these new spectral regions (e.g. the thulium-doped fibre amplifier or TDFA for S-band) are less mature and less efficient compared to the erbium-doped fibre amplifier (EDFA) used in the conventional C-and L-bands. Indeed, there is little published research into the system-level power consumption considerations for these new wavelength bands.

Distributed Raman amplification (DRA) offers an alternative amplification scheme for UWB systems by using the fibre as the gain medium. While DRA allows for multi-band amplification with reduced ASE noise compared to lumped amplifiers, it is much less power efficient \cite{lundberg2017power}. While considerable analysis exists for C-band systems, no power efficiency analysis of DRA or hybrid-amplification has yet been conducted in a multi-band scenario.

In this work, we study the power consumption of hybrid-amplified SCL-band links. We measure the electrical-to-optical power conversion efficiency of a commercial TDFA, C-band EDFA and L-band EDFA. We use the Gaussian noise (GN) model with inter-channel stimulated Raman scattering (ISRS) and DRA to investigate the power efficiency (energy per bit) of both CL and SCL-band transmission scenarios with lumped amplification and hybrid pump-optimised backward-pumped DRA.


\section{Power conversion efficiency measurements}

Assuming uniform power per WDM channel, the total amplifier power consumption of a hybrid-amplified SCL-band link can be described by \cite{lundberg2017power}:
\begin{align} \label{eq:totalpow}
    P_{\text{tot}} = N_{\text{span}} \Bigg( \sum_{ i \in \{S,C,L\} } \frac{1}{\eta_{i}} &N_{\text{ch}} P_{\text{ch}} \bigg( 1 - \frac{1}{G_{i}} \bigg)\nonumber \\+ P_{\text{mm}}^{i}
    + &\frac{1}{\eta_{\text{R}}} \sum_{k =1}^{N_{R}}  P_{\text{R},k} \Bigg)
\end{align}


\noindent where $N_{\text{span}}$ is the span number, $\eta_i$ is the electrical-to-optical (wallplug) power conversion efficiency (PCE) in band $i$, $N_{\text{ch}}$ is the number of channels in band $i$, $P_{\text{ch}}$ is the per-channel launch power, $G_i$~is the lumped amplifier gain for band $i$, $P_{\text{mm}}$ is the lumped amplifier monitoring and management power consumption, $N_{\text{R}}$ is the number of Raman pumps and $P_{\text{R}}$ is the output power per Raman pump. The first and second terms reflect the power consumption of the SCL lumped amplifiers and the third accounts for the Raman pumps.

While EDFA power conversion efficiency has been studied extensively \cite{frisch2013electrical,desbruslais2015maximizing,Liang2021}, little analysis is available for the more recent TDFA. Determining $\eta_i$ in Eq.~\ref{eq:totalpow} is possible analytically, but depends on several factors including: pump wavelengths and powers, pump PCE, doped fibre length and conversion efficiency, gain-flattening filters, thermo-electric cooling (TEC) efficiency etc. These parameters depend heavily on link design and are not always available in commercial specifications. Here, we directly measure the power draw of our commercial amplifiers at the wall socket using off-the-shelf power monitors to determine our amplifier wallplug efficiencies from Eq. \ref{eq:totalpow}. All amplifiers and pumps measured here were used in previous SCL-band transmission experiments \cite{yang2024experimental, yang2025transmission, Hazarika2024}.

\input{figure/pce_combined}

We fully loaded the entire S-, C- and L-bands with spectrally-flattened ASE and measured the wallplug power consumption and output power as pump drive current was increased. The input power of each band was fixed at 2~dBm. The wallplug efficiency is shown in Fig.~\ref{fig:pce}(a) for three amplifiers. The Amonics C- and L-band benchtop EDFAs have 23~dBm saturation and the Fiberlabs TDFA has 20.5~dBm saturation. The wallplug efficiency does not include $P_{\text{mm}}$ which was found to be 8~W, similar to previous estimations in literature of 10~W \cite{lundberg2017power}. The C-band EDFA PCE of 5\% is also similar to that in prior studies \cite{desbruslais2015maximizing,lundberg2017power}. The L-band EDFA PCE is lower than the C-band due to lower gain efficiency of EDFAs at long wavelengths, requiring more pump power. The TDFA PCE is 1\% as it uses three pumps instead of one. As no deployed TDFA system currently exists, this result is used as an estimate for deployed S-band systems.

We carried out a power consumption measurement for our Raman pumps, shown in Fig.~\ref{fig:pce}(b). The pump sources are Anritsu laser diodes centred at 1365~nm, 1385~nm, 1405~nm and 1425~nm. The power consumption of these pumps are practically identical, thus only one curve is shown. Here the output power only accounts for the Raman pump, not the signal power after DRA. The Raman PCE depends on both fibre and signal properties, thus we report the pump power draw only.

\section{GN model with ISRS}

To estimate power consumption in a hybrid-SCL link, we used the ISRS GN model in the presence of DRA with the modulation format correction to calculate backward-pumped DRA gain and SNR \cite{buglia2024closed,buglia2024ultra,buglia2023modulationformat}. We modelled a fully-loaded system with 140~GBd dual-polarisation 64-QAM channels with a 150~GHz spacing and transceiver SNR of 20~dB. Two scenarios were considered: (i) CL-band (1530nm-1620nm, 73 channels) and (ii) SCL-band (1460nm-1620nm, 127 channels). The launch power is uniform with 2~dBm per channel. The fibre is 80~km SMF-28 with dispersion 16.5~ps/(nm$\cdot$km), nonlinear coefficient of 1.13 W$^{-1}$km$^{-1}$ and wavelength-dependent attenuation of 0.2~dB/km at 1550~nm\cite{buglia2024ultra}. The lumped S-, C- and L-band amplifiers have noise figure of 6, 5 and 6~dB respectively. For backward-pumped DRA, a particle swarm optimiser was used to determine optimum pump wavelengths and powers (up to 250~mW per pump) to maximise the Shannon capacity \cite{buglia2024ultra}. We use the results from Fig.~\ref{fig:pce} alongside Eq. \ref{eq:totalpow} to obtain the lumped and DRA power consumption (assuming lumped output power is saturated). Since DRA is a multi-band amplifier, we use the ratio of DRA gain in each band to determine the per-band DRA power consumption.

\section{Power evaluation results}

\input{figure/powcon1span}

The power consumption of each band is shown in Fig. \ref{fig:powcon} for 1 span. $P_{\text{mm}}$ was set to zero to investigate the power consumption without overheads. From Eq. \ref{eq:totalpow}, the total power consumption scales linearly with $N_{\text{span}}$, thus the ratios do not change with distance. Without DRA, the S-band consumes more than double the C+L bands combined due to the low TDFA PCE. As more DRA pumps are introduced, the S-band benefits most from DRA gain than the C/L bands, with DRA power consumption quickly dominating. A similar trend is visible in the CL scenario but with similar DRA consumption between C+L bands.

Fig.~\ref{fig:throughput} shows the throughput per band $C_{\text{th}}$ for different amplification schemes for CL and SCL systems. At 1 span, the throughput is limited by transceiver noise, thus DRA offers little gain in both systems (except in the S-band, where DRA is the most beneficial) \cite{buglia2022launch}. The S-band provides the largest capacity as it has the largest number of channels. Despite this, at 100 spans the lumped S-band underperforms both C- and L-bands since ISRS and amplifier ASE noise limit the throughput. Consequently, DRA provides the most benefit to the S-band, corresponding to the high DRA power consumption shown in Fig.~\ref{fig:powcon}. In the CL system, the L-band achieves the highest throughput as it benefits the most from ISRS and DRA.

\input{figure/throughput}

\input{figure/jouleperbit2by3_v2}

Fig.~\ref{fig:jouleperbit2x3} shows the energy per bit for each scenario for 1 and 100 spans and varying $P_{\text{mm}}$. Overall, the S-band has the highest energy per bit which increases with Raman pump count. However, as $P_{\text{mm}}$ increases, its contribution dominates $P_{\text{tot}}$ in Eq. \ref{eq:totalpow} and the energy per bit becomes proportional to $1/C_{\text{th}}$, thus the S-band energy per bit approaches the C-band. For 100 spans, hybrid-DRA energy per bit decreases as more pumps are added, with S-band (total) energy per bit reducing by 17\%, 25\% and 36\% (12\%, 20\% and 26\%) compared to lumped amplification for $P_{\text{mm}} =$ 0~W, 2~W and 8~W respectively, where 8~W corresponds to our benchtop amplifier measurements. Comparing Figs. \ref{fig:pce} and \ref{fig:throughput} at 100 spans, the DRA-enhanced throughput in the S-band offsets the increased DRA power consumption, leading to improved overall energy per bit for SCL systems. Since transceiver throughput adaptively depends on link quality, the throughput enhancement by hybrid-DRA increases spectral efficiency, potentially leading to additional improvements in end-to-end power efficiency per bit.


\section{Conclusion}

We investigated the power consumption of hybrid-amplified SCL-band links. We found that the low power conversion efficiency of the TDFA contributes more than double the power consumption than C- and L-band lumped amplifiers alone. Hybrid-DRA not only leads to higher throughput than lumped amplification, but can lead to improvements in energy per bit in long-haul systems, potentially translating to lower power consumption of pluggable coherent transceivers.


\section{Acknowledgements}
This work was supported by EPSRC Grant EP/R035342/1 Transforming Networks - building an intelligent optical infrastructure (TRANSNET), EP/W015714/1 Extremely Wideband Optical Fibre Communication Systems (EWOC), and EP/V007734/1 EPSRC Strategic equipment grant. Polina Bayvel is supported under a Royal Society Research Professorship.

\printbibliography
\vspace{-4mm}

\end{document}

%% file: figure/pce_combined.tex
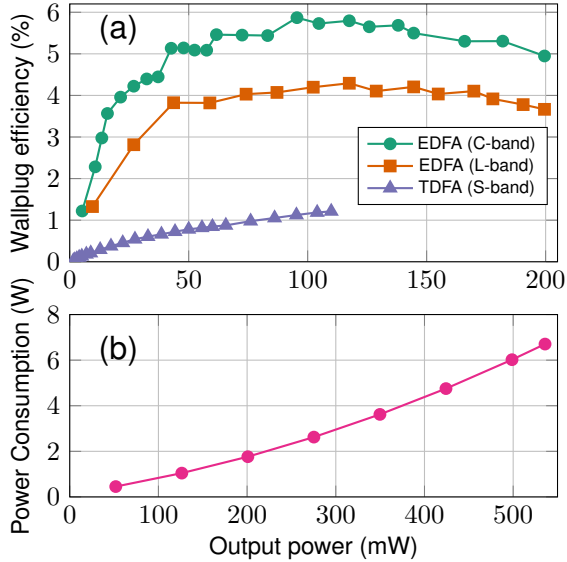
\begin{figure}[h]

\begin{tikzpicture}

    \begin{groupplot}
    [
    legend columns=1,
    width=0.5\textwidth,height=5cm,
    grid=both,
    legend style={fill opacity=1, draw opacity=1, text opacity=1, at={(0.98,0.22)}, anchor=south east, draw=black, nodes={scale=0.65, transform shape}},
    ylabel near ticks,
    xlabel near ticks,
    ylabel shift = -2 pt,
    xlabel shift = -2 pt,
    clip marker paths=true,
    group style={group size=1 by 2, horizontal sep=0.8cm,  vertical sep=0.7cm,xlabels at=edge bottom,ylabels at=edge left},
    label style={font=\small}
    ]
    
    \nextgroupplot[
    ylabel=Wallplug efficiency (\%),
    ytick distance=1,
    ymin=0,
    ymax=6.2,
    xmin=0,
    xmax=205,
    ]

        \addplot[green1,thick,mark=*] table[x=xc,y=yc] {data/pcedatac.tsv};
        \addlegendentry{EDFA (C-band)}
        
         \addplot[orange1,thick,mark=square*] table[x=xl,y=yl] {data/pcedatal.tsv};
        \addlegendentry{EDFA (L-band)}

        \addplot[purple1,thick,mark=triangle*,mark size=2.5] table[x=xs,y=ys] {data/pcedatas.tsv};
        \addlegendentry{TDFA (S-band)}

        \node[at={(axis description cs:0.1,0.9)},font=\large]{(a)};

    \nextgroupplot[
    grid=both,
    width=0.5\textwidth,height=4cm,
    ylabel=Power Consumption (W),
    ymin=0,
    ymax=8,
    xmin=0,
    xmax=550,
    clip marker paths=true,
    xlabel=Output power (mW),
    legend style={fill opacity=1, draw opacity=1, text opacity=1, at={(0.3,0.7)}, anchor=south west, draw=black, nodes={scale=1, transform shape}},
    ]
     \addplot[pink1,thick,mark=*] table[x=x,y=y] {data/ramanpumpdata.tsv};

     \node[at={(axis description cs:0.1,0.8)},font=\large]{(b)};
         
    \end{groupplot}

\end{tikzpicture}

\caption{(a) Measured electrical-to-optical wallplug power conversion efficiency versus output power for SCL-band amplifiers. (b) Measured wall-plug power consumption versus output power for a 1365~nm Raman pump.}
\label{fig:pce}

\end{figure}

%% file: figure/powcon1span.tex
\pgfplotsset{
        compat=1.9,
        compat/bar nodes=1.8,
    }

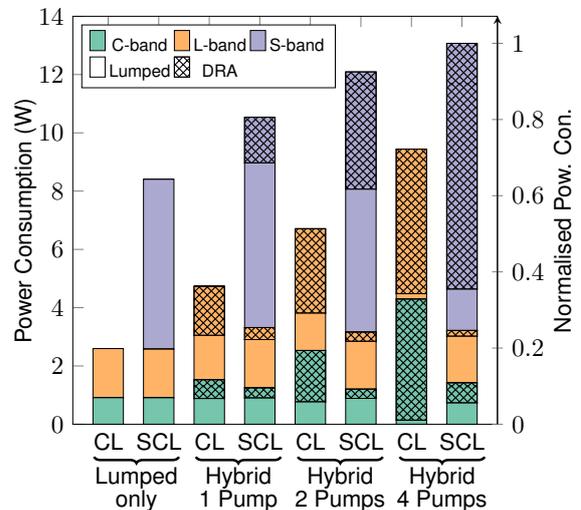
\begin{figure}[b!]
\centering

\begin{tikzpicture}[scale=0.95]
\begin{axis}[
    legend style={fill opacity=1, draw opacity=1, text opacity=1, at={(0.02,0.98)}, anchor=north west, draw=black, nodes={scale=0.67, transform shape}},
    legend columns = 3,
    ybar stacked,
	bar width=12pt,
    x = 0.7cm,
    xtick=data,
    ylabel shift = -6 pt,
    xticklabels={CL,SCL,CL,SCL,CL,SCL,CL,SCL},
    xticklabel style={text width=2cm,align=center},
    ylabel={Power Consumption (W)},
    ymin = 0, ymax = 14,
    ytick distance = 2,
    ylabel near ticks,
    xlabel near ticks,
    ytick pos=left,
    clip mode=individual,
    xticklabel style={text width=2cm,align=center,font=\small},
    ylabel style={font=\small},
    ]

\addplot [black,fill=green1!60] table [y=clumped] {data/powconseg.tsv};

 \addplot [black, fill=green1!60, postaction={pattern=crosshatch}] table [y=craman] {data/powconseg.tsv};

 \addplot [black,fill=orange!60] table [y=llumped] {data/powconseg.tsv};

  \addplot [black, fill=orange!60, postaction={pattern=crosshatch}] table [y=lraman] {data/powconseg.tsv};

 \addplot [black,fill=purple1!60] table [y=slumped] {data/powconseg.tsv};

  \addplot [black, fill=purple1!60, postaction={pattern=crosshatch}] table [y=sraman] {data/powconseg.tsv};

\addlegendimage{fill=white}
\addlegendimage{fill=white, postaction={pattern=crosshatch} }

\legend{C-band,,L-band,,S-band,, Lumped, DRA}

\node[at={(axis description cs:0.15,-0.16)},font=\small,align=center]{Lumped\\[-0.4ex]only};
\draw [decorate,decoration = {brace,mirror},line width = 1] (axis description cs:0.05,-0.08)--(axis description cs:0.25,-0.08);

    \node[at={(axis description cs:0.39,-0.16)},font=\small,align=center]{Hybrid\\[-0.4ex]1 Pump};
\draw [decorate,decoration = {brace,mirror},line width = 1] (axis description cs:0.29,-0.08)--(axis description cs:0.49,-0.08);

    \node[at={(axis description cs:0.63,-0.16)},font=\small,align=center]{Hybrid\\[-0.4ex]2 Pumps};
\draw [decorate,decoration = {brace,mirror},line width = 1] (axis description cs:0.53,-0.08)--(axis description cs:0.73,-0.08);

    \node[at={(axis description cs:0.87,-0.16)},font=\small,align=center]{Hybrid\\[-0.4ex]4 Pumps};
\draw [decorate,decoration = {brace,mirror},line width = 1] (axis description cs:0.77,-0.08)--(axis description cs:0.97,-0.08);

\end{axis}

\begin{axis}[
    x = 0.817cm,
    ymin = 0, ymax = 14/13.0715,
    ytick distance = 0.2,
        ylabel shift = -3 pt,
      hide x axis,
      axis y line=right,
      ylabel={Normalised Pow. Con.},
      ylabel near ticks,
      ylabel style={font=\small},
      bar width=12pt,
    ]

    \addplot [black, draw=none] table [y=sramanrel] {data/powconseg.tsv};
    
    \end{axis}

\end{tikzpicture}

\caption{Power consumption of CL and SCL amplification after 1 span. Solid: lumped, crosshatch: DRA.}
\label{fig:powcon}

\end{figure}

%% file: figure/throughput.tex
\begin{figure*}[ht]

\begin{tikzpicture}

    \begin{groupplot}
    [
    legend columns=2,
    width=0.38\textwidth,height=5cm,
    grid=both,
    legend style={fill opacity=1, draw opacity=1, text opacity=1, at={(0.5,0.2)}, anchor=south, draw=black, nodes={scale=0.7, transform shape}},
    ylabel near ticks,
    xlabel near ticks,
    ylabel shift = -2 pt,
    xlabel shift = -2 pt,
    title style={at={(0.5,0.94)},font=\bfseries\footnotesize},
    clip marker paths=true,
    group style={group size=3 by 1, horizontal sep=0.65cm,  vertical sep=0.7cm,xlabels at=edge bottom,ylabels at=edge left},
    label style={font=\small},
    ]
    
    \nextgroupplot[
    ylabel=Throughput per band $C_{\text{th}}$ (Tb/s),
    minor y tick num = 0,
    ymin=50,
    ymax=85,
    xtick = data,
    xticklabels={Lumped\\only,Hybrid\\1 Pump,Hybrid\\2 Pumps,Hybrid\\4 Pumps},
    xticklabel style={text width=2cm,align=center,font=\footnotesize},
    title = {1 span, 80 km},
    ]

        \addplot[green1,thick,dashed,no marks, forget plot] table[y=CL1C] {data/throughput.tsv};
        \addplot[green1,mark=*,only marks, draw=black] table[y=CL1C] {data/throughput.tsv};
        
        \addlegendimage{gray1, dashed}
        \addplot[orange1,thick,dashed,no marks, forget plot] table[y=CL1L] {data/throughput.tsv};
        \addplot[orange1,mark=square*,only marks,draw=black] table[y=CL1L] {data/throughput.tsv};

        \addlegendimage{gray1}

        \addplot[purple1,thick,no marks,forget plot] table[y=SCL1S] {data/throughput.tsv};
        \addplot[purple1,mark=triangle*,only marks,draw=black,mark size = 2.9pt] table[y=SCL1S] {data/throughput.tsv};
        
        \addplot[green1,thick,no marks, forget plot] table[y=SCL1C] {data/throughput.tsv};
        \addplot[green1,mark=*,only marks, draw=black] table[y=SCL1C] {data/throughput.tsv};

        \addplot[orange1,thick,no marks, forget plot] table[y=SCL1L] {data/throughput.tsv};
        \addplot[orange1,mark=square*,only marks,draw=black] table[y=SCL1L] {data/throughput.tsv};

        \legend{C-Band, CL system,L-Band, SCL system,S-Band}

    \nextgroupplot[
    ymin=40,
    ymax=72,
    xtick = data,
    xticklabels={Lumped\\only,Hybrid\\1 Pump,Hybrid\\2 Pumps,Hybrid\\4 Pumps},
    xticklabel style={text width=2cm,align=center,font=\footnotesize},
    legend style={at={(0.4,0.35)}},
    title = {10 spans, 800 km},
    ]






        \addplot[green1,thick,dashed,no marks] table[y=CL10C] {data/throughput.tsv};
        \addplot[green1,mark=*,only marks, draw=black] table[y=CL10C] {data/throughput.tsv};

        \addplot[orange1,thick,dashed,no marks] table[y=CL10L] {data/throughput.tsv};
        \addplot[orange1,mark=square*,only marks,draw=black] table[y=CL10L] {data/throughput.tsv};

        \addplot[green1,thick,no marks] table[y=SCL10C] {data/throughput.tsv};
        \addplot[green1,mark=*,only marks, draw=black] table[y=SCL10C] {data/throughput.tsv};

        \addplot[orange1,thick,no marks] table[y=SCL10L] {data/throughput.tsv};
        \addplot[orange1,mark=square*,only marks,draw=black] table[y=SCL10L] {data/throughput.tsv};

        \addplot[purple1,thick,no marks] table[y=SCL10S] {data/throughput.tsv};
        \addplot[purple1,mark=triangle*,only marks,draw=black,mark size = 2.9pt] table[y=SCL10S] {data/throughput.tsv};


    \nextgroupplot[
    ymin=18,
    ymax=45,
    xtick = data,
    xticklabels={Lumped\\only,Hybrid\\1 Pump,Hybrid\\2 Pumps,Hybrid\\4 Pumps},
    xticklabel style={text width=2cm,align=center,font=\footnotesize},
    legend style={at={(0.68,0.03)}},
    title = {100 spans, 8000 km},
    ]
        \addplot[green1,thick,dashed,no marks] table[y=CL100C] {data/throughput.tsv};
        \addplot[green1,mark=*,only marks, draw=black] table[y=CL100C] {data/throughput.tsv};

        \addplot[orange1,thick,dashed,no marks] table[y=CL100L] {data/throughput.tsv};
        \addplot[orange1,mark=square*,only marks,draw=black] table[y=CL100L] {data/throughput.tsv};

        \addplot[green1,thick,no marks] table[y=SCL100C] {data/throughput.tsv};
        \addplot[green1,mark=*,only marks, draw=black] table[y=SCL100C] {data/throughput.tsv};

        \addplot[orange1,thick,no marks] table[y=SCL100L] {data/throughput.tsv};
        \addplot[orange1,mark=square*,only marks,draw=black] table[y=SCL100L] {data/throughput.tsv};

        \addplot[purple1,thick,no marks] table[y=SCL100S] {data/throughput.tsv};
        \addplot[purple1,mark=triangle*,only marks,draw=black,mark size = 2.9pt] table[y=SCL100S] {data/throughput.tsv};


    \end{groupplot}

\end{tikzpicture}

\caption{Throughput for different amplification schemes after 1, 10 and 100 spans. Dashed: CL system, solid: SCL system.}
\label{fig:throughput}

\end{figure*}
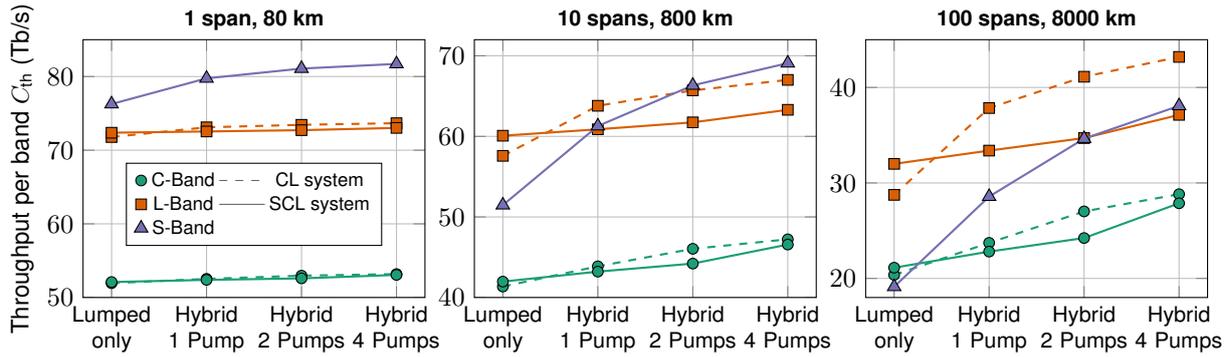

%% file: figure/jouleperbit2by3_v2.tex
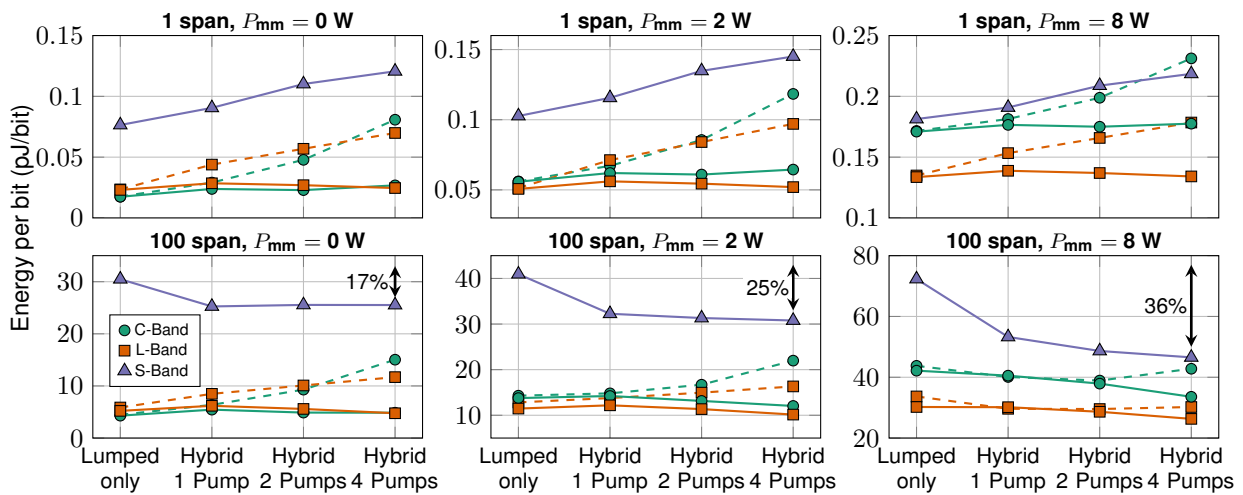
\begin{figure*}[hb]

\begin{tikzpicture}

    \begin{groupplot}
    [
    legend columns=1,
    width=0.37\textwidth,height=4cm,
    grid=both,
    legend style={fill opacity=1, draw opacity=1, text opacity=1, at={(0.5,0.3)}, anchor=south west, draw=black, nodes={scale=0.6, transform shape}},
    ylabel near ticks,
    xlabel near ticks,
    ylabel shift = -2 pt,
    xlabel shift = -2 pt,
    title style={at={(0.5,0.88)},font=\bfseries\footnotesize},
    clip marker paths=true,
    group style={group size=3 by 3, horizontal sep=0.9cm,  vertical sep=0.5cm,xlabels at=edge bottom,ylabels at=edge left},
    label style={font=\small},
    ]
    
    \nextgroupplot[
    ylabel = Energy per bit (pJ/bit),
    ymin=0,
    ymax=0.15,
    xticklabels=\none,
    legend style={at={(0.1,0.72)}},
    title = {1 span, $P_{\text{mm}} =$ 0 W},
    scaled ticks=false,
     yticklabel style={/pgf/number format/.cd,fixed,precision=3},
     y label style={at={(axis description cs:-.15,0)}},
    ]

        \addplot[green1,thick,dashed,no marks] table[y=CL1C] {data/jouleperbitpmm0.tsv};
        \addplot[green1,mark=*,only marks, draw=black] table[y=CL1C] {data/jouleperbitpmm0.tsv};

        \addplot[orange1,thick,dashed,no marks] table[y=CL1L] {data/jouleperbitpmm0.tsv};
        \addplot[orange1,mark=square*,only marks,draw=black] table[y=CL1L] {data/jouleperbitpmm0.tsv};

        \addplot[green1,thick,no marks] table[y=SCL1C] {data/jouleperbitpmm0.tsv};
        \addplot[green1,mark=*,only marks, draw=black] table[y=SCL1C] {data/jouleperbitpmm0.tsv};

        \addplot[orange1,thick,no marks] table[y=SCL1L] {data/jouleperbitpmm0.tsv};
        \addplot[orange1,mark=square*,only marks,draw=black] table[y=SCL1L] {data/jouleperbitpmm0.tsv};

        \addplot[purple1,thick,no marks] table[y=SCL1S] {data/jouleperbitpmm0.tsv};
        \addplot[purple1,mark=triangle*,only marks,draw=black,mark size = 2.9pt] table[y=SCL1S] {data/jouleperbitpmm0.tsv};


    \nextgroupplot[
    minor y tick num = 0,
    ymin=0.03,
    ymax=0.16,
    xticklabels=\none,
    legend style={at={(0.1,0.72)}},
    yticklabel style={/pgf/number format/.cd,fixed,precision=3},
    title = {1 span, $P_{\text{mm}} =$ 2 W},
    ]

        \addplot[green1,thick,dashed,no marks] table[y=CL1C] {data/jouleperbitpmm2.tsv};
        \addplot[green1,mark=*,only marks, draw=black] table[y=CL1C] {data/jouleperbitpmm2.tsv};

        \addplot[orange1,thick,dashed,no marks] table[y=CL1L] {data/jouleperbitpmm2.tsv};
        \addplot[orange1,mark=square*,only marks,draw=black] table[y=CL1L] {data/jouleperbitpmm2.tsv};

        \addplot[green1,thick,no marks] table[y=SCL1C] {data/jouleperbitpmm2.tsv};
        \addplot[green1,mark=*,only marks, draw=black] table[y=SCL1C] {data/jouleperbitpmm2.tsv};

        \addplot[orange1,thick,no marks] table[y=SCL1L] {data/jouleperbitpmm2.tsv};
        \addplot[orange1,mark=square*,only marks,draw=black] table[y=SCL1L] {data/jouleperbitpmm2.tsv};

        \addplot[purple1,thick,no marks] table[y=SCL1S] {data/jouleperbitpmm2.tsv};
        \addplot[purple1,mark=triangle*,only marks,draw=black,mark size = 2.9pt] table[y=SCL1S] {data/jouleperbitpmm2.tsv};


    \nextgroupplot[
    minor y tick num = 0,
    ymin=0.1,
    ymax=0.25,
    xticklabels=\none,
    legend style={at={(0.1,0.72)}},
    title = {1 span, $P_{\text{mm}} =$ 8 W},
    ]

        \addplot[green1,thick,dashed,no marks] table[y=CL1C] {data/jouleperbitpmm8.tsv};
        \addplot[green1,mark=*,only marks, draw=black] table[y=CL1C] {data/jouleperbitpmm8.tsv};

        \addplot[orange1,thick,dashed,no marks] table[y=CL1L] {data/jouleperbitpmm8.tsv};
        \addplot[orange1,mark=square*,only marks,draw=black] table[y=CL1L] {data/jouleperbitpmm8.tsv};

        \addplot[green1,thick,no marks] table[y=SCL1C] {data/jouleperbitpmm8.tsv};
        \addplot[green1,mark=*,only marks, draw=black] table[y=SCL1C] {data/jouleperbitpmm8.tsv};

        \addplot[orange1,thick,no marks] table[y=SCL1L] {data/jouleperbitpmm8.tsv};
        \addplot[orange1,mark=square*,only marks,draw=black] table[y=SCL1L] {data/jouleperbitpmm8.tsv};

        \addplot[purple1,thick,no marks] table[y=SCL1S] {data/jouleperbitpmm8.tsv};
        \addplot[purple1,mark=triangle*,only marks,draw=black,mark size = 2.9pt] table[y=SCL1S] {data/jouleperbitpmm8.tsv};

        
        %
        %
        %
        %
        %
        %
        %
        %
        %

     \nextgroupplot[
    minor y tick num = 0,
    ymin=0,
    ymax=35,
    xtick = data,
    xticklabels={Lumped\\only,Hybrid\\1 Pump,Hybrid\\2 Pumps,Hybrid\\4 Pumps},
    xticklabel style={text width=2cm,align=center,font=\footnotesize},
    legend style={at={(0.05,0.28)}},
    title = {100 span, $P_{\text{mm}} =$ 0 W},
    ]

        \addplot[green1,thick,dashed,no marks] table[y=CL100C] {data/jouleperbitpmm0.tsv};
        \addplot[green1,mark=*,only marks, draw=black] table[y=CL100C] {data/jouleperbitpmm0.tsv};

        \addplot[orange1,thick,dashed,no marks] table[y=CL100L] {data/jouleperbitpmm0.tsv};
        \addplot[orange1,mark=square*,only marks,draw=black] table[y=CL100L] {data/jouleperbitpmm0.tsv};

        \addplot[green1,thick,no marks] table[y=SCL100C] {data/jouleperbitpmm0.tsv};
        \addplot[green1,mark=*,only marks, draw=black] table[y=SCL100C] {data/jouleperbitpmm0.tsv};

        \addplot[orange1,thick,no marks] table[y=SCL100L] {data/jouleperbitpmm0.tsv};
        \addplot[orange1,mark=square*,only marks,draw=black] table[y=SCL100L] {data/jouleperbitpmm0.tsv};

        \addplot[purple1,thick,no marks] table[y=SCL100S] {data/jouleperbitpmm0.tsv};
        \addplot[purple1,mark=triangle*,only marks,draw=black,mark size = 2.9pt] table[y=SCL100S] {data/jouleperbitpmm0.tsv};

        \legend{,,,,,C-Band,,L-Band,,S-Band}

        \draw [stealth-stealth,line width = 1] (axis cs:4,33)--(axis cs:4,27) node[midway,fill=none,inner sep=2pt,font=\footnotesize,anchor=east,] {17\%};

    \nextgroupplot[
    minor y tick num = 0,
    ymin=5,
    ymax=45,
    xtick = data,
    xticklabels={Lumped\\only,Hybrid\\1 Pump,Hybrid\\2 Pumps,Hybrid\\4 Pumps},
    xticklabel style={text width=2cm,align=center,font=\footnotesize},
    legend style={at={(0.65,0.55)}},
    title = {100 span, $P_{\text{mm}} =$ 2 W},
    ]

        \addplot[green1,thick,dashed,no marks] table[y=CL100C] {data/jouleperbitpmm2.tsv};
        \addplot[green1,mark=*,only marks, draw=black] table[y=CL100C] {data/jouleperbitpmm2.tsv};

        \addplot[orange1,thick,dashed,no marks] table[y=CL100L] {data/jouleperbitpmm2.tsv};
        \addplot[orange1,mark=square*,only marks,draw=black] table[y=CL100L] {data/jouleperbitpmm2.tsv};

        \addplot[green1,thick,no marks] table[y=SCL100C] {data/jouleperbitpmm2.tsv};
        \addplot[green1,mark=*,only marks, draw=black] table[y=SCL100C] {data/jouleperbitpmm2.tsv};

        \addplot[orange1,thick,no marks] table[y=SCL100L] {data/jouleperbitpmm2.tsv};
        \addplot[orange1,mark=square*,only marks,draw=black] table[y=SCL100L] {data/jouleperbitpmm2.tsv};

        \addplot[purple1,thick,no marks] table[y=SCL100S] {data/jouleperbitpmm2.tsv};
        \addplot[purple1,mark=triangle*,only marks,draw=black,mark size = 2.9pt] table[y=SCL100S] {data/jouleperbitpmm2.tsv};


        \draw [stealth-stealth,line width = 1] (axis cs:4,43)--(axis cs:4,33) node[midway,fill=white,inner sep=1pt,font=\footnotesize,anchor=east] {25\%};

    \nextgroupplot[
    minor y tick num = 0,
    ymin=20,
    ymax=80,
    xtick = data,
    xticklabels={Lumped\\only,Hybrid\\1 Pump,Hybrid\\2 Pumps,Hybrid\\4 Pumps},
    xticklabel style={text width=2cm,align=center,font=\footnotesize},
    legend style={at={(0.65,0.55)}},
    title = {100 span, $P_{\text{mm}} =$ 8 W},
    ]

        \addplot[green1,thick,dashed,no marks] table[y=CL100C] {data/jouleperbitpmm8.tsv};
        \addplot[green1,mark=*,only marks, draw=black] table[y=CL100C] {data/jouleperbitpmm8.tsv};

        \addplot[orange1,thick,dashed,no marks] table[y=CL100L] {data/jouleperbitpmm8.tsv};
        \addplot[orange1,mark=square*,only marks,draw=black] table[y=CL100L] {data/jouleperbitpmm8.tsv};

        \addplot[green1,thick,no marks] table[y=SCL100C] {data/jouleperbitpmm8.tsv};
        \addplot[green1,mark=*,only marks, draw=black] table[y=SCL100C] {data/jouleperbitpmm8.tsv};

        \addplot[orange1,thick,no marks] table[y=SCL100L] {data/jouleperbitpmm8.tsv};
        \addplot[orange1,mark=square*,only marks,draw=black] table[y=SCL100L] {data/jouleperbitpmm8.tsv};

        \addplot[purple1,thick,no marks] table[y=SCL100S] {data/jouleperbitpmm8.tsv};
        \addplot[purple1,mark=triangle*,only marks,draw=black,mark size = 2.9pt] table[y=SCL100S] {data/jouleperbitpmm8.tsv};



        \draw [stealth-stealth,line width = 1] (axis cs:4,77)--(axis cs:4,50) node[midway,fill=white,inner sep=1pt,font=\footnotesize,anchor=east] {36\%};
         
    \end{groupplot}

\end{tikzpicture}

\caption{Energy per bit for various amplification schemes for different $P_{\text{mm}}$ at 1 and 100 spans. Dashed: CL, solid: SCL system.}
\label{fig:jouleperbit2x3}

\end{figure*}